\documentclass[11pt]{article}
\usepackage[]{amsmath}
\usepackage{amscd}
\begin{document}
\title{Nonlinear superposition formula for $N=1$ \\ supersymmetric KdV Equation}
\author{Q. P. Liu and Y. F. Xie \\[2mm]
Department of Mathematics,\\
China University of Mining and Technology,\\
 Beijing 100083, China.}
\date{}
\maketitle
\begin{abstract}
In this paper, we derive a B\"{a}cklund transformation for the
supersymmetric Korteweg-de Vries equation. We also construct a
nonlinear superposition formula, which allows us to rebuild
systematically for the supersymmetric KdV equation the soliton
solutions of Carstea, Ramani and Grammaticos.
\end{abstract}

 The celebrated Korteweg-de Vries (KdV) equation was extended into super framework
by Kupershmidt \cite{kup1} in 1984. Shortly afterwards, Manin and
Radul \cite{mr} proposed another super KdV system which is a
particular reduction of their general supersymmetric
Kadomstev-Petviashvili hierarchy. In \cite{ma}, Mathieu pointed
out that the super version of Manin and Radul for the KdV equation
is indeed invariant under a space supersymmetric transformation,
while Kupershmidt's version does not. Thus, the Manin-Radul's
super KdV is referred to the supersymmetric KdV equation.

We notice that the supersymmetric KdV equation has been studied
extensively in literature and a number of interesting properties
has been established. We mention here the infinite conservation
laws \cite{ma}, bi-Hamiltonian structures \cite{op}, bilinear form
\cite{my}\cite{crg},
 Darboux transformation \cite{liu-mm}.

By the constructed Darboux transformation, Ma\~{n}as and one of us
calculated the soliton solutions for the supersymmetric KdV
system. This sort of solutions was also obtained by Carstea in the
framework of bilinear formalism \cite{ca}. However, these
solutions are characterized by the presentation of some constraint
on soliton parameters. Recently, using super-bilinear operators,
Carstea, Ramani and Grammaticos \cite{crg} constructed explicitly
new two- and three-solitons for the supersymmetric KdV equation.
These soliton solutions are interesting since they are free of any
constraint on soliton parameters. Furthermore, the fermionic part
of these solutions is dressed through the interactions.

In addition to the  bilinear form approach, B\"{a}cklund
transformation (BT) is also a powerful method to construct
solutions. Therefore, it is interesting to see if the soliton
solutions of Carstea-Ramani-Grammaticos can be constructed by BT
approach. In this paper, we first construct a BT for the
supersymmetric KdV equation. Then, we derive a nonlinear
superposition formula. In this way, the soliton solutions can be
produced {\em systematically}. We explicitly show that the
two-soliton solution of Carstea, Ramani and Grammaticos appears
naturally in the framework of BT.

To introduce the supersymmetric extension for the KdV equation, we
recall some terminology and notations. The classical spacetime is
$(x,t)$ and we extend it to a super-spacetime $(x,t,\theta)$,
where $\theta$ is a Grassmann  odd variable. The dependent
variable $u(x,t)$ in the KdV equation is replaced by a fermionic
variable $\Phi=\Phi(x,t,\theta)$. Now the supersymmetric KdV
equation reads as
\begin{equation}
\Phi_t-3(\Phi {\cal D}\Phi)_x+\Phi_{xxx}=0 \label{skdv},
\end{equation}
where ${\cal D}={\partial\over \partial\theta}+\theta{\partial\over\partial x}$ is the superderivative.
Mathieu found the following supersymmetric version of Gardner type map
\begin{equation}\label{gardner-m}
\Phi=\chi+\epsilon \chi_x+\epsilon^2\chi({\cal D}\chi),
\end{equation}
where $\epsilon$ is an ordinary (bosonic) parameter.

It is easy to show that $\chi$ satisfies the following supersymmetric Gardner equation
\begin{equation}\label{gardner-e}
\chi_t-3(\chi{\cal D }\chi)_x-3\epsilon^2({\cal D}\chi)(\chi{\cal
D }\chi)_x+\chi_{xxx}=0.
\end{equation}

This map was used in \cite{ma} to prove that there exists an
infinite number of conservation laws for the supersymmetric KdV
equation (\ref{skdv}). In the classical case, Gardner type of map
was studied extensively by Kupershmidt \cite{kup}. It is well
known that such map may be used to construct interesting BT.  We
will show that it is also the case for the supersymmetric KdV
equation.

We notice that the supersymmetric Gardner equation
(\ref{gardner-e}) is invariant under $\epsilon\rightarrow
-\epsilon$. The new solution of the supersymmetric KdV equation
corresponding to with $-\epsilon$ is denoted as $\tilde{\Phi}$.
Thus we have
\begin{equation}\label{gardner-et}
\tilde{\Phi}=\chi-\epsilon \chi_x+\epsilon^2\chi({\cal D}\chi).
\end{equation}
From above relations (\ref{gardner-e}-\ref{gardner-et}), we find
\begin{eqnarray}
\Phi-\tilde{\Phi}&=&2\epsilon \chi_x,\label{e1}\\
\Phi+\tilde{\Phi}&=&2\chi+2\epsilon\chi({\cal D\chi}).\label{e2}
\end{eqnarray}
Let us introduce the potentials as follows
\[
\Phi=\Psi_x, \quad \tilde{\Phi}=\tilde{\Psi}_x,
\]
thus, the equation (\ref{e1}) provides us
\begin{equation}
\chi={1\over 2\epsilon}(\Psi-\tilde{\Psi}), \label{e3}
\end{equation}
Eliminating $\chi$  between the equations (\ref{e2}) and
(\ref{e3}), we arrive at a BT
\begin{equation}\label{bt}
(\Psi+\tilde{\Psi})_x=\lambda
(\Psi-\tilde\Psi)+\frac{1}{2}(\Psi-\tilde\Psi)({\cal D}\Psi-{\cal
D}\tilde\Psi),
\end{equation}
where $\lambda={1/\epsilon}$ is the B\"{a}cklund parameter.

The transformation (\ref{bt}) is in fact the spatial part of BT.
Its temporal counterpart can be easily worked out. We also remark
here that the BT is reduced to the well known BT for the classical
KdV equation if $\eta=0$, as it should be.

A BT can be used to generate special solutions. If we start with
the trivial solution $\tilde{\Psi}=0$, we obtain
\begin{equation}\label{seeds}
\Psi=-\frac{2a(\zeta+\theta\lambda)e^{\lambda x-\lambda^3
t}}{1+ae^{\lambda x-\lambda^3 t}},
\end{equation}
where $a$ is an Grassmann even constant and $\zeta$ Grassmann odd
one. When $a>0$, we have $1$-soliton solution for the
supersymmetric KdV equation (\ref{skdv}). For $a<0$, we have a
singular solution. (cf. \cite{we} for the classical KdV equation)

 Next, we shall work out a superposition
formula. To this end, we start with the seed solution $\Psi_0$.
After one step transformation, our seed is transformed to $\Psi_1$
with parameter $\lambda_1$, to $\Psi_2$ with parameter
$\lambda_2$, respectively. Then we do second step transformation:
starting  with $\Psi_1$, we obtain $\Psi_{12}$ with $\lambda_2$
while starting with $\Psi_2$, we obtain $\Psi_{21}$ with
$\lambda_1$. By Bianchi's theorem of permutability, one should
have $\Psi_{12}=\Psi_{21}$. For convenience, we denote
\[
\Psi_3=:\Psi_{12}=\Psi_{21}.
\]
\begin{center}
\setlength{\unitlength}{1.3cm}
\begin{picture}(7,1.5)
\thicklines \put(1.3,-0.2){\vector(2,-1){2}}
\put(1.3,0.2){\vector(2,1){2}} \put(0.9,0){$\Psi_0$}
\put(3.5,1.2){$\Psi_1$} \put(4.2,1.2){\vector(2,-1){2}}
\put(4.2,-1.2){\vector(2,1){2}} \put(3.5,-1.4){$\Psi_2$}
\put(6.3,0){$\Psi_{3}$} \put(1.8, 0.8){$\lambda_1$}
\put(1.9,-0.8){$\lambda_2$}
\put(5,0.9){$\lambda_2$}\put(5,-1.1){$\lambda_1$}
\put(2.5,-2.2){Bianchi's diagram}
\end{picture}
\end{center}

\vspace{3cm}

We list down the relations obtained
\begin{equation}\label{e4}
(\Psi_0+\Psi_1)_x=\lambda_1(\Psi_1-\Psi_0)+{1\over
2}(\Psi_1-\Psi_0)({\cal D}\Psi_1-{\cal D}\Psi_0),
\end{equation}
and
\begin{equation}\label{e5}
(\Psi_0+\Psi_2)_x=\lambda_2(\Psi_2-\Psi_0)+{1\over
2}(\Psi_2-\Psi_0)({\cal D}\Psi_2-{\cal D}\Psi_0),
\end{equation}
and
\begin{equation}\label{e6}
(\Psi_1+\Psi_3)_x=\lambda_2(\Psi_3-\Psi_1)+{1\over
2}(\Psi_3-\Psi_1)({\cal D}\Psi_3-{\cal D}\Psi_1),
\end{equation}
and
\begin{equation}\label{e7}
(\Psi_2+\Psi_3)_x=\lambda_1(\Psi_3-\Psi_2)+{1\over
2}(\Psi_3-\Psi_2)({\cal D}\Psi_3-{\cal D}\Psi_2).
\end{equation}
Subtraction (\ref{e4}) form (\ref{e5}), we have
\begin{eqnarray}\label{e8}
(\Psi_2-\Psi_1)_x&=&\lambda_2\Psi_2-\lambda_1\Psi_1+(\lambda_1-\lambda_2)\Psi_0+{1\over
2}\Psi_2({\cal D}\Psi_2)-{1\over 2}\Psi_2({\cal
D}\Psi_0)\nonumber\\
&&-{1\over 2}\Psi_0({\cal D}\Psi_2)-{1\over 2}\Psi_1({\cal
D}\Psi_1)+{1\over 2}\Psi_1({\cal D}\Psi_0)+{1\over 2}\Psi_0({\cal
D}\Psi_1),
\end{eqnarray}
similarly, from (\ref{e6}) and (\ref{e7}) we have
\begin{eqnarray}\label{e9}
(\Psi_2-\Psi_1)_x&=&\left[\lambda_1-\lambda_2+{1\over 2}({\cal
D}\Psi_1)-{1\over 2}({\cal D}\Psi_2)\right]\Psi_3+{1\over
2}(\Psi_1-\Psi_2)({\cal
D}\Psi_3)\nonumber\\
&&+\lambda_2\Psi_1-\lambda_1\Psi_2+{1\over 2}\Psi_2({\cal
D}\Psi_2)-{1\over 2}\Psi_1({\cal D}\Psi_1).\label{16}
\end{eqnarray}

Now equations (\ref{e8}) and (\ref{e9}) give us
\begin{eqnarray}
\lefteqn{\left[\lambda_1-\lambda_2+{1\over 2}({\cal
D}\Psi_1)-{1\over 2}({\cal
D}\Psi_2)\right](\Psi_3-\Psi_0)}\nonumber\\
&&+{1\over 2}(\Psi_1-\Psi_2)\left[({\cal
D}\Psi_3)+2\lambda_1+2\lambda_2-({\cal D}\Psi_0
)\right]=0.\label{17}
\end{eqnarray}
The equation (\ref{17}) is a differential equation for $\Psi_3$.
Solving it we obtain
\begin{equation}\label{nsf}
\Psi_3=\Psi_0-\frac{(\lambda_1+\lambda_2)(\Psi_1-\Psi_2)}{\lambda_1-\lambda_2+({\cal
D}\Psi_1)-({\cal D}\Psi_2) },
\end{equation}
this is the superposition formula for supersymmetric KdV equation
(\ref{skdv}).
 Substituting $\Psi_i=\eta_i+\theta v_i$ ($i=0, 1, 2, 3$) into (\ref{nsf}), our superposition
 formula may be decomposed into
\begin{eqnarray*}
v_3&=&v_0-\frac{(\lambda_1+\lambda_2)(v_1-v_2)}{\lambda_1-\lambda_2+v_1-v_2}-\frac{(\lambda_1+\lambda_2)(\eta_1-\eta_2)(\eta_{1,x}-\eta_{2,x})}{
(\lambda_1-\lambda_2+v_1-v_2)^2},\\
\eta_3&=&\eta_0-\frac{(\lambda_1+\lambda_2)(\eta_1-\eta_2)}{
(\lambda_1-\lambda_2+v_1-v_2)}.
\end{eqnarray*}
It is clear that
our nonlinear superposition formula reduces to the well-known
superposition formula for the KdV as it should be. The advantage
to have a superposition formula is that it is an algebraic one and
can be used easily to find solutions.

Using the solutions (\ref{seeds}) as our seeds, we may construct a
2-soliton solution of (\ref{skdv}) by means of our superposition
formula. Indeed,
 let
\[
\Psi_0=0,\quad
\Psi_1=-\frac{2a_1(\zeta_1+\theta\lambda_1)e^{\lambda_1
x-\lambda_1^3 t}}{1+a_1e^{\lambda_1 x-\lambda_1^3 t}},\quad
\Psi_2=-\frac{2a_2(\zeta_2+\theta\lambda_2)e^{\lambda_2
x-\lambda_2^3 t}}{1+a_2e^{\lambda_2 x-\lambda_2^3 t}}
\]
then from our nonlinear superposition formula (\ref{nsf}), we
obtain
\[
\Psi_3=\frac{2\left[\zeta_1a_1e^{\delta_1}-\zeta_2a_2e^{\delta_2}+(\zeta_1-\zeta_2)a_1a_2e^{\delta_1+\delta_2}+
\theta(\lambda_1a_1e^{\delta_1}-\lambda_2a_2e^{\delta_2}+(\lambda_1-\lambda_2)a_1a_2e^{\delta_1+\delta_2})\right]}
{(\lambda_1+\lambda_2)^{-1}\left[(\lambda_1-\lambda_2)-(\lambda_1+\lambda_2)a_1e^{\delta_1}+(\lambda_1+\lambda_2)
a_2e^{\delta_2}-(\lambda_1-\lambda_2)a_1a_2e^{\delta_1+\delta_2}\right]}
\]
where $\delta_i=\lambda_ix-\lambda_i^3t+\theta\zeta_i$ ($i=1,2$).
Now, taking
\[
\lambda_1>\lambda_2, \quad a_1=-{\lambda_1-\lambda_2\over
\lambda_1+\lambda_2}, \qquad a_2={\lambda_1-\lambda_2\over
\lambda_1+\lambda_2}
\]
we recover the 2-soliton solution found first by Carstea, Ramani
and Grammaticos \cite{crg}. As in the classical KdV case
\cite{we}, we generate this 2-soliton from a regular solution and
a singular solution.

We could continue this process to build the higher soliton
solutions and  the calculation will be tedious but
straightforward.

\bigskip

 {\bf Acknowledgment} The work is supported in part by National
Natural Scientific Foundation of China (grant number 10231050) and
Ministry of Education of China.

\end{document}